# Deep Convolutional Neural Networks to Predict Mutual Coupling Effects in Metasurfaces

*Sensong An[1], Bowen Zheng[1], Mikhail Y. Shalaginov[2], Hong Tang[1], Hang Li[1], Li Zhou[1], Yunxi Dong[1], Mohammad Haerinia[1], Anuradha Murthy Agarwal[2,3], Clara Rivero-Baleine[4], Myungkoo Kang[5], Kathleen A. Richardson[5], Tian Gu[2], Juejun Hu[2], Clayton Fowler[1,\*], Hualiang Zhang[1,\*]*

[1]Department of Electrical & Computer Engineering, University of Massachusetts Lowell, Lowell, Massachusetts 01854, USA

[2]Department of Materials Science & Engineering, Massachusetts Institute of Technology, Cambridge, Massachusetts 02139, USA

[3]Materials Research Laboratory, Massachusetts Institute of Technology, Cambridge, Massachusetts 02139, USA

[4]Lockheed Martin Corporation, Orlando, Florida 32819, USA

[5]CREOL, University of Central Florida, Orlando, Florida 32816, USA

Abstract: Metasurfaces have provided a novel and promising platform for the realization of compact and large-scale optical devices. The conventional metasurface design approach assumes periodic boundary conditions for each element, which is inaccurate in most cases since the near-field coupling effects between elements will change when surrounded by non-identical structures. In this paper, we propose a deep learning approach to predict the actual electromagnetic (EM) responses of each target meta-atom placed in a large array with near-field coupling effects taken into account. The predicting neural network takes the physical specifications of the target meta-atom and its neighbors as input, and calculates its phase and amplitude in milliseconds. This approach can be applied to explain metasurfaces' performance deterioration caused by mutual coupling and further used to optimize their efficiencies once combined with optimization algorithms. To demonstrate the efficacy of this methodology, we obtain large improvements in efficiency for a beam deflector and a metalens over the conventional design approach. Moreover, we show the correlations between a metasurface's performance and its design errors caused by mutual coupling are not bound to certain specifications (materials, shapes, etc.). As such, we envision that this approach can be readily applied to explore the mutual coupling effects and improve the performance of various metasurface designs.

Metamaterials, along with their two dimensional (2D) versions, metasurfaces, have attracted wide attentions in recent years due to their unique low profile and lightweight properties as compared to their conventional bulk optics counterparts. Meta-atoms, the building blocks of metasurfaces, constructed with either all-dielectric[1-3] or plasmonic[4-6] nanoresonators, are capable of achieving engineered phase and amplitude control at the element level and thus enable accurate wave front control with subwavelength resolution. The most widely adopted metasurface design approach includes two steps: 1) calculate the amplitude and phase masks necessary for desired functionalities, fitted to square or hexagonal grids, and 2) find meta-atoms with performance closest to the target of each grid for the final design. Accurate and efficient meta-atom on-demand design approaches remain the main challenge with metasurface designs.

To design meta-atoms with maximum efficiency and accurate phase gradients, a common method is to consider structures with simple geometric shapes (such as circles[7, 8], rectangles[9, 10], H-shapes[11, 12] and plasmonic thin layers[13, 14]) and perform a parameter sweep over all dimensions to assemble a library for the full design space. Then best-fit meta-atoms are selected from the library to most closely resemble the ideal amplitude/phase map. Beyond this brute force approach, numerous optimization algorithms[15-17], deep neural networks (DNN)[18-24] and DNN-optimization adjoint methods[25-27] have also been proposed recently for fast inverse design of meta-atoms with complicated shapes. In these meta-atom design approaches, unit cell boundary conditions were adopted during full wave simulations, which assumes that each meta-atom structure under consideration is part of an infinite 2D array of identical structures. Thus, the amplitude and phase response calculations of the meta-atoms are based on the assumption that near-field coupling perturbations originate from identical neighbors. However, in real metasurface designs, each meta-atom is usually surrounded by non-identical meta-atoms, for which near-field coupling effects will differ from those used to calculate the original response. As a result, the phase and amplitude of each meta-atom will be perturbed from their predicted values, and so this method is accurate only when the mutual coupling effects between each meta-atom and its various neighbors are identical, which will generally not be the case. We illustrate near-field mutual coupling effects between several types of meta-atoms, including circles[7, 8], rectangles[9] and ring-shaped thin layers[6] in Fig. 1. By comparing the near-field distributions between electric field simulations of meta-atoms with identical neighbors and differing neighbors, considerable deviations inside the center meta-atom can be observed in all four cases.

In general, it is noted that the meta-atoms structures that utilize multipole resonances to achieve phase and amplitude manipulation (Huygens surface) and meta-atoms with lower refractive indexes are particularly prone to the mutual coupling effect. Instances have been reported where the mutual coupling effect plays such an important role in a metasurfaces' overall performance that measures must be taken to minimize its effect. In Ref.[8] a periodic arrangement of 2 by 2 meta-atoms was adopted in order to decrease the coupling effects in a metasurface hologram, while in Ref.[28] the metasurface hologram was divided into several 17.35 by 17.35 $\mu m^2$ sub-arrays each constructed with identical meta-atoms. In Ref.[29] the height and diameters of the cylindrical meta-atoms that formed the metasurface beam deflector were slightly adjusted to achieve higher efficiency. In Ref.[30] a genetic algorithm (GA) was employed to find the meta-atoms' optimal dimensions and positions with strong coupling between neighbors taken into consideration. In some other works, the mutual coupling effects between adjacent meta-atoms were not only investigated, but also utilized to enhance the metasurfaces' performance. In Ref. [31] a strongly coupled resonator design was proposed and demonstrated in the terahertz and optical regimes, in which the coupling between neighboring resonators was tuned to enhance the effective refractive index. In Ref. [32] the near-field effects in high-index Mie-resonant nanoparticles were studied, and the distances between neighboring meta-atoms were tuned to realize continuous relative phase changes. In Ref. [33] a numerical method (so-called local phase method (LPM)) was proposed to obtain the phase of each meta-atom within the metasurfaces while considering the mutual coupling effects. This approach quantifies the phase error of each element inside the metasurfaces, which enables the optimization of metasurfaces on an element level while accounting for near-field coupling effects. While it provides a way to measure the meta-atoms' accurate phase responses, the target meta-atom and all its neighbors need to be simulated as a whole in order to derive the performance of a single meta-atom, which is computationally intensive and time consuming.

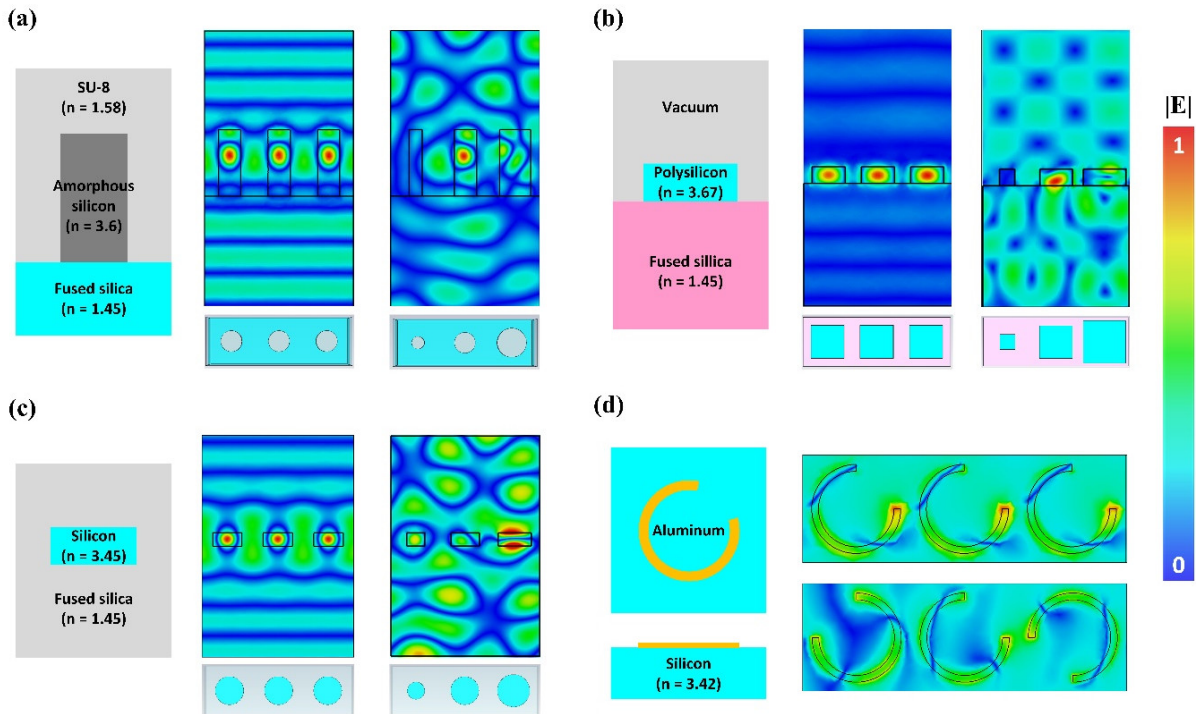

**Fig. 1. Simulated near-field electric field distribution of meta-atoms placed in identical and different neighbors.** Structures and materials of each meta-atom model are showing on the left in each subplot. The meta-atom in the center of each 3 meta-atoms are placed among identical (middle) and different (right) neighbors. In each case, the perturbing effects of near-field coupling are apparent from the significant change in the field distribution inside the central meta-atom when surrounded by non-identical neighbors. Simulations are performed using full-wave simulation tool CST, following the simulation setups in **(a)** Ref. [7], **(b)** Ref. [9], **(c)** Ref. [8] and **(d)** Ref. [6].

In this paper, we propose a DNN approach to efficiently predict the meta-atoms' real phase and amplitude responses while accounting for the influences of its neighbors. When fully trained, the DNN is able to predict the perturbed EM response of a meta-atom given the dimensions of itself and all its neighbors within a 2 by 2 square wavelength area. Importantly, the accurate forward predictions can be achieved in milliseconds, which enables the fast optimization of various metasurface devices, including beam deflectors, lenses and holograms that are composed of densely arranged meta-atoms prone to amplitude drop or phase error caused by mutual coupling. To demonstrate the efficacy of this DNN approach, we employed the fully trained DNN to optimize several beam deflectors and focusing lenses, whose efficiencies have been greatly improved according to the full-wave simulation results.

**Data collection.** Without loss of generality, we started with the transmissive all-dielectric metasurface design presented in Ref. [9], which is constructed with rectangular-shaped high-index (n = 3.67) polysilicon nanoblocks sitting on a low-index (n = 1.45) fused silica substrate (Fig. 2a). The wavelength of interest is set to be 1.55 μm, with the lattice constant and the height of the nanoblocks fixed to be 800 nm and 270 nm, respectively. By carefully selecting the length and width of each nanoblock, this meta-atom design can cover full 2π phase while maintaining high transmission (Fig. 2b), which is essential for most optical applications.

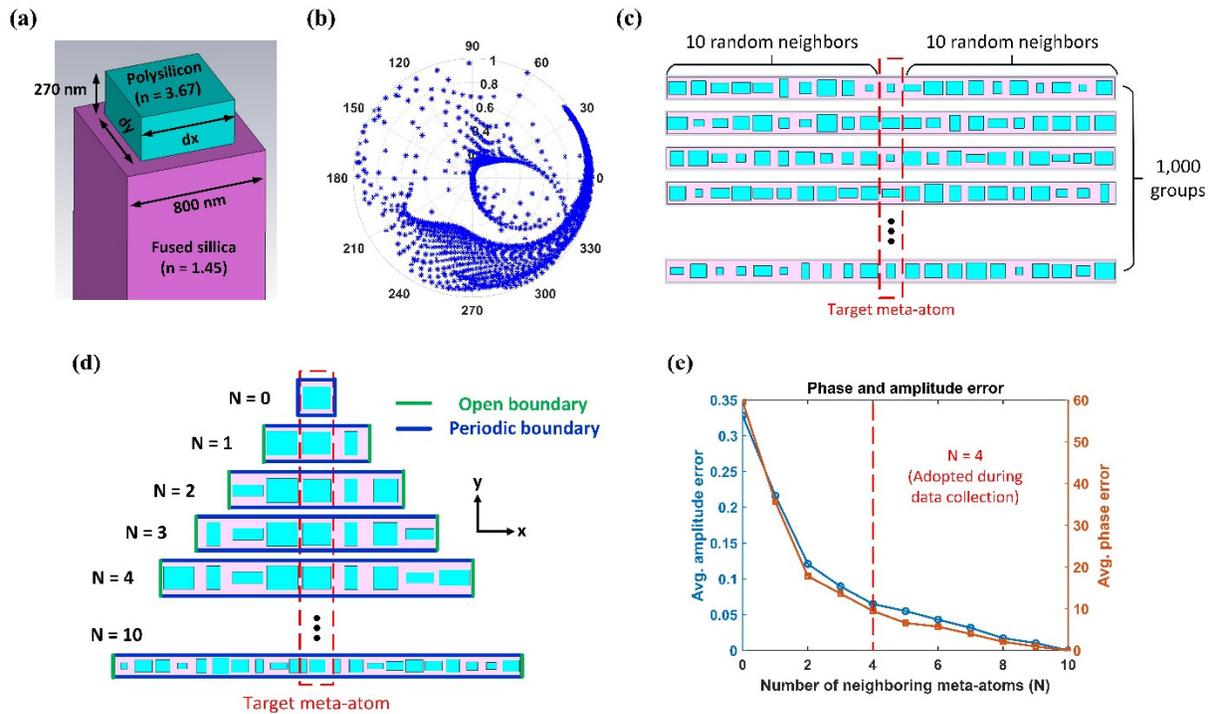

**Fig. 2. Data collection setup. (a)** Schematic of a meta-atom composed of a silicon nanoblock sitting on top of bulk-fused silica substrate. **(b)** Transmittance phase and amplitude responses of meta-atoms with varying length, *dx,* and width, *dy,* at 1.55 μm wavelength. **(c)** 1,000 randomly generated target meta-atoms (circled in red) surrounded by 10 random neighbors on each side. **(d)** Different numbers of neighboring meta-atoms (N) were considered in the simulation to derive accurate EM responses of each target meta-atom. **(e)** The average amplitude and phase error as a function of N. N = 0 represents the results derived with periodic boundaries. N = 4 was adopted during the data collection process.

First, the actual phase and amplitude responses of randomly-generated meta-atoms placed among different random neighbors (hereinafter referred to as "local responses") were calculated to assemble the training dataset. Simulation of each group of data starts by modeling the whole structure, which includes the target meta-atom and its neighbors, using the commercial software CST. The equivalent sources ($J_s, M_s$) of the target meta-atom were then obtained (by the Field Source Monitor in CST) and exported to a new simulation file to measure its local phase and amplitude responses. For simplicity, the problem was limited to one dimension (1D), meaning only coupling effects along the x-axis (as shown in Fig. 2) were considered, and periodic boundary conditions were set along the y-axis. Considering that mutual coupling effects decrease with distance, it is necessary to determine the number of neighboring meta-atoms that needed to be considered in each simulation. To do so, we randomly generated a group of meta-atoms that consisted of a target meta-atom surrounded by 10 differing meta-atoms on each side (Fig. 2c). We initially simulated the structure with no differing neighbors considered (periodic boundary conditions) and gradually increased the number of neighbors accounted for in the simulation (Fig. 2d) to examine how the local response of the target meta-atom changes. We treat the local response with all 10 neighbors on each side as a reference point with zero error, because improvements in accuracy are marginal with the inclusion of additional neighbors beyond this point. We repeated this experiment on 1,000 different groups of meta-atoms and calculated the average phase and amplitude error. As shown in Fig. 2e, the simulation results of target meta-atoms with a periodic boundary (hereinafter referred to as "periodic responses") have significant amplitude and phase mean absolute error (MAE) of 0.33 and 59.14 degrees compared to when the number of neighboring meta-atoms (N) is 10. As the number of neighboring meta-atoms increases, the error gradually decreases. The large error values in Fig. 2e are caused by the abrupt changes in phase gradients that can occur in randomly arranged meta-atoms. According to the results, even considering only the single nearest meta-atom on each side of the target would largely improve its EM response's accuracy, which is in accordance with our predictions. To strike a balance between simulation accuracy and optimization difficulty (as well as data collection costs), we selected N = 4 during the data collection process. The final dataset was composed of the physical dimensions and local responses of over 100,000 randomly-generated meta-atoms that were each surrounded by 4 randomly-generated meta-atoms on each side. The lengths (*dx*) and widths (*dy*) of all nanoblocks were within the range of [0.25 μm, 0.75 μm] with a minimum step of 6.25 nm.

**Network training and results.** In order to realize the fast predictions of meta-atoms' local responses under various boundary conditions, a predicting neural network (PNN) was constructed based on a convolutional neural network (CNN) architecture. The PNN takes the dimensions of the target meta-atom and dimensions of its neighbors as input, and generates the prediction of target meta-atoms' local responses quickly. As shown in Fig. 3, 2D cross sections of the target meta-device were extracted and pixelated into an 1152 by 128 figure. The figure is binarized such that "1"s represent dielectric regions and "0"s represent voids. This figure was then processed through 6 consecutive convolutional layers, during which hidden features such as relationships between the nanoblock's dimensions and its local responses, as well as the impact of its non-identical neighbors, were extracted and calculated. After flattening the output of the CNNs into a one dimensional vector and passing through 3 fully connected layers, the prediction results for the complex transmission coefficients of the target meta-atom were generated and ready for evaluation. Throughout the network, a ReLU activation function was applied to each layer except for the last one, for which there was no activation function.

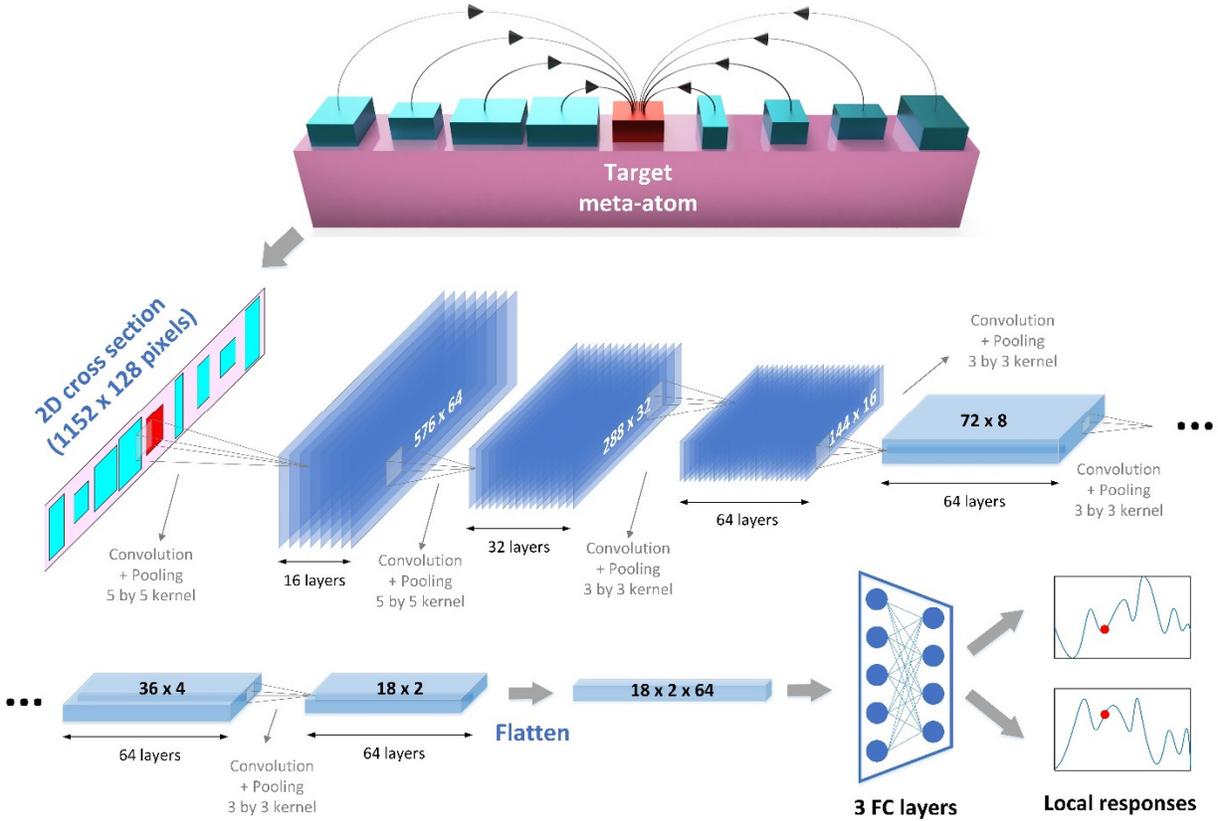

**Fig. 3. Network architecture.** 2D cross sections of the target meta-atom (in red) and its neighbors (in blue) were processed through 6 consecutive convolution and pooling layers, then flattened into a 1D vector (1 by 2304). After being processed with 3 more fully connected layers (containing 512, 64 and 2 nodes, respectively), the real and imaginary parts of the transmission coefficients of the target meta-atom were ready for evaluation.

Over 100,000 groups of the collected training data were randomly split into a training set and a test set, containing 70% and 30% of the total training data, respectively. The test set was used to evaluate the trained network's performance on data that was not used during training. During training, the PNN-predicted local responses were compared with the labels (accurate local responses) to calculate the mean square error (MSE), which was minimized by inversely tuning the parameters in the hidden neural layers. When the training was completed, the average MSE was $9e^{-6}$ for the training set and $7e^{-5}$ for the test set, respectively (Fig. 4a), which corresponds to an average prediction standard deviation of 0.005 for amplitude and 3.15 degrees for phase at the target wavelength for each target meta-atom. We demonstrate the accuracy of the well-trained PNN with several samples that were randomly selected from the test set. As shown in Fig. 4b, six target meta-atoms (marked with red) surrounded by 4 different meta-atoms on each side (marked in yellow) were set as input of the fully-trained PNN. The PNN predicted local responses are labeled with blue stars, while the accurate results calculated with CST are labeled in red. Good agreements have been achieved between the predicted and accurate results, while significant differences between the local responses and periodic responses can also be observed, which demonstrates the detriment of mutual coupling effects.

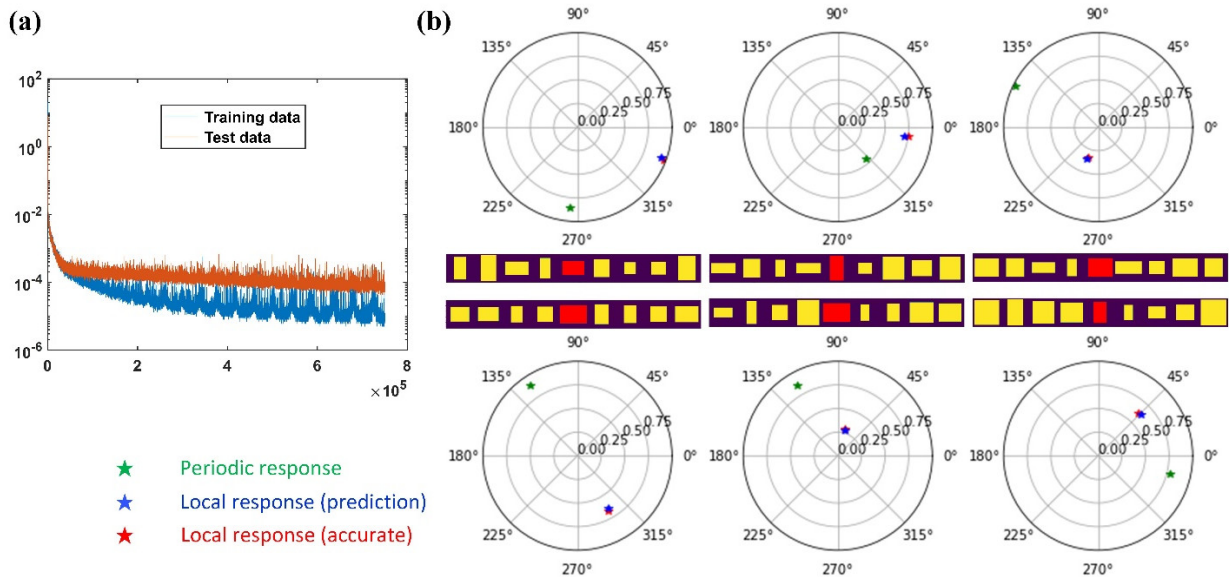

**Fig. 4. Learning curve and results. (a)** Average MSE of the training set (red curve) and test set (blue curve) during the training process. **(b)** Six groups of results randomly selected from the test set and demonstrated. The target meta-atoms are marked in red, while the neighbors are marked in yellow. Periodic responses are labeled with green stars, while local responses predicted with PNN and calculated with CST are labeled in blue and red stars, respectively.

**Applications.** With the help of this fully trained PNN, the actual responses of the target meta-atoms (when deployed in a meta-device) can be predicted in milliseconds. This efficient and accurate tool enables the fast optimization of metasurface devices that suffer from performance deterioration caused by mutual coupling. As shown in Fig. 5a, we combined the PNN with a global optimization algorithm to inversely optimize the performance of metasurfaces. First, a conventional metasurface was designed based on the periodic boundary condition assumption and assigned as input. The optimization algorithm generates new designs based on the input, then each meta-atom inside the new design will be evaluated by the PNN to find out its local phase and amplitude responses. Subsequently, dimensions of each element will be further tuned to minimize the difference between the current result and the design goal, and a new design will be given by the optimization algorithm. The optimization process stops when the stopping criteria is met or the maximum number of iterations is reached. In this case, a Dual Sequence Simulated Annealing Algorithm was adopted as the optimizer. We started with the optimization of a beam deflector and compared the performance of the optimized design with the initial design to demonstrate the efficacy of this optimization approach.

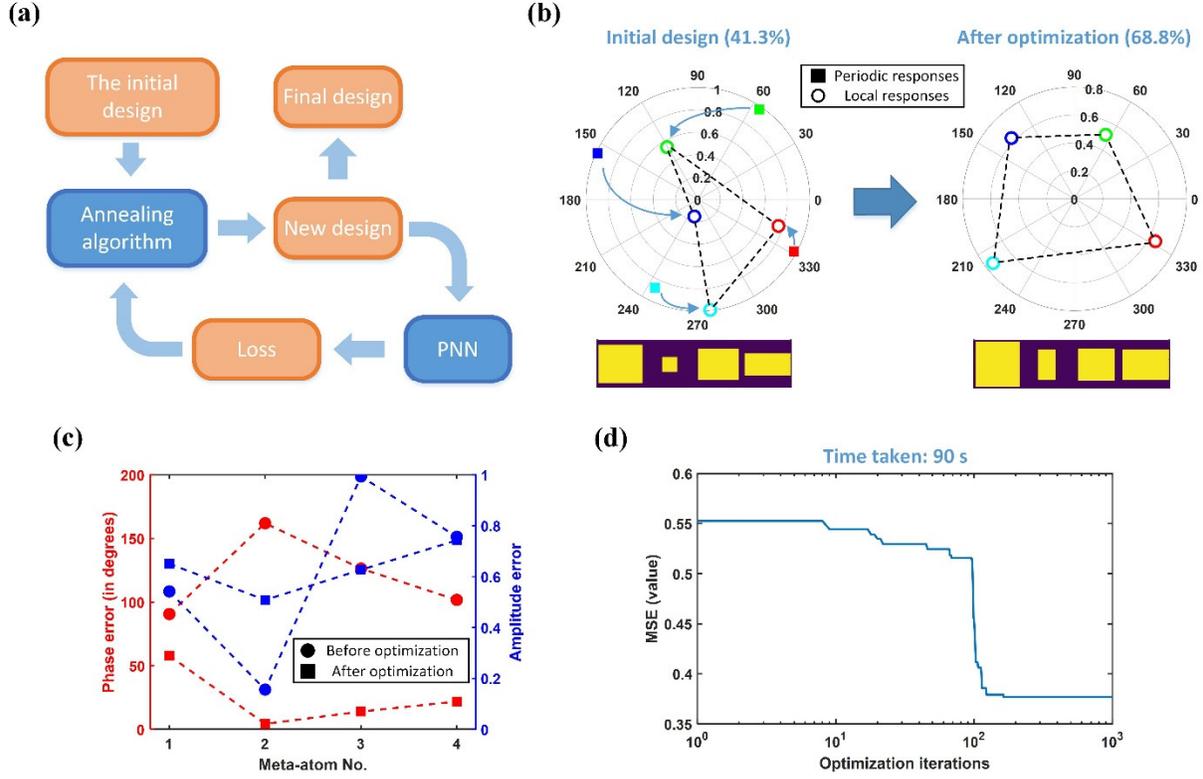

**Fig. 5. Optimization of a beam deflector composed of 1 by 4 meta-atom elements. (a)** Flow chart of the combined optimization approach. **(b)** Periodic responses and local responses of the initial design and the final optimized design. Top view of these two designs are shown on the bottom. Dielectric nanoblocks are yellow, while dark color represents the void. **(c)** Phase error and amplitude of the optimized design compared to the initial design. **(d)** Value of the objective function during the optimization.

Fig. 5b presents the optimization result of a beam deflector composed of 1 by 4 meta-atom elements using the proposed optimization approach. For the initial design shown in the left, periodic responses of each meta-atom are marked with squares in the polar plot, while their corresponding local responses (i.e. actual responses) are marked with empty circles. The phase targets are set to 60, 150, 240 and 330 degrees, with the amplitude targets set to 1. Although the periodic responses are nearly perfect, with almost unity transmission and precise 90-degree phase shift between each pair of adjacent meta-atoms, their local responses deviate from the design goals due to the mutual coupling effects, resulting in a relatively low efficiency (41.3%). We then started the optimization with this initial design as a starting point and minimized the objective function, defined as:

$$E = \frac{1}{N} \sum_{i=1}^{N} \sqrt{((\phi_i - \widehat{\phi}_i)/\pi)^2 + (A_i - \widehat{A}_i)^2} \qquad (1)$$

where N is the number of the meta-atoms in the device, $\phi_i$ and $A_i$ are the local phase and amplitude responses, $\widehat{\phi}_i$ and $\widehat{A}_i$ are the target phase and amplitude responses, respectively. After 1,000 iterations of optimization, the local responses of the optimized meta-atoms as shown on the right (Fig. 5b) are much closer to the design goals, which also leads to a higher efficiency of 68.8%, meaning nearly 70% of the incident energy is deflected to the 1st order. Comparisons of the phase error and amplitude error between the initial design and the optimized design are presented in Fig. 5c. The values of the objective function during the optimization process are shown in Fig. 5d, after which the average phase error has been reduced from 30.3° (initial design) to 10.4° (optimized design) while the average amplitude remains roughly the same, indicating the efficacy of the proposed approach.

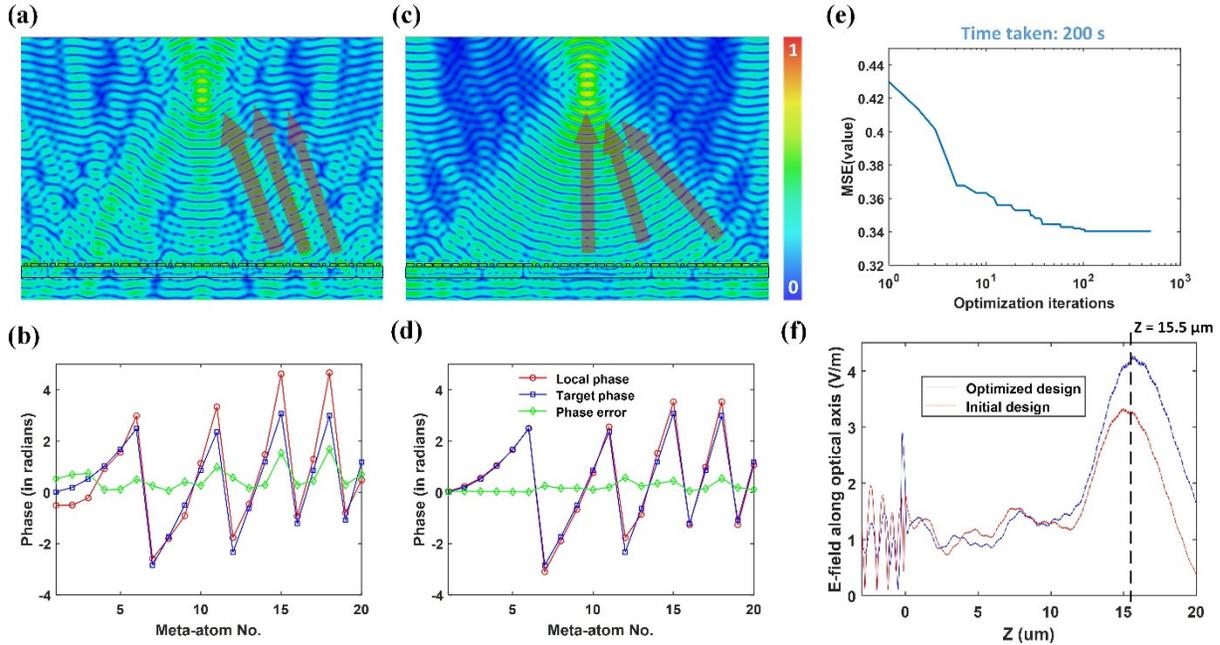

**Fig. 6. Optimization of a high numerical aperture metalens composed of 40 meta-atoms. (a)** Simulated electric field (real part) of the initial design, with target phase (blue), local phase (red) and phase error (green) of each meta-atom showing in **(b)**. **(c)** Simulated electric field of the optimized design, with target phase, local phase and phase error showing in **(d)**. **(e)** Value of the objective function during the optimization. **(f)** Simulated electric field along the optical (Z) axis, with an x-polarized incidence for the initial design (red) and the optimized design (blue).

The advantage of this combined optimization approach becomes even more evident when applied to larger size metasurfaces, such as high numerical aperture (NA) metalenses. Here we designed a one dimensional (1D) metalens composed of 40 meta-atoms with a focal length of 10 wavelengths (15.5 μm) using the conventional approach. The real part of the transmitted electric field of this lens as simulated with CST is shown in Fig. 6a. Large discontinuities of the wave front can be observed, as well as unfocused transmitted energy that points in various directions (marked with red arrows). The local phase responses of the meta-atoms in the initial design are shown in Fig. 6b, where the phase error of each meta-atom is also calculated and plotted. Due to the symmetrical nature of the metalens, we only listed the responses of the meta-atoms in the right half. The phase errors between the targets and the local responses are almost 100 degrees at certain points, which leads to the phase discontinuities. Subsequently, this initial design was set as a starting point for the optimization using the proposed network and the same objective function in equation (1) was adopted. After 500 iterations of optimization (finished in 200 seconds), the local phase responses of the optimized meta-atoms (Fig. 6c) are much closer to the targets, with the average phase error reduced from 30.4° to 10.4° (Fig. 6d). Value of the objective function during the optimization were plotted in Fig. 6e. As a result, the electric field at focal spot (z = 15.5 µm) increased from 3.24 V/m to 4.17 V/m, equivalent to a 28.7% and 65.6% enhancement in electric field intensity and power intensity, respectively (Fig. 6e). This clearly indicates that the transmitted energy is more confined to the focal spot due to the more accurate phase front. To further demonstrate that this approach is compatible with full size metasurface designs, a high NA metalens composed of 200 meta-atoms was also designed and optimized. Focal length of this larger metalens is set to be 50 wavelengths (77.5 µm). Similarly, after 500 iterations of optimization (finished in 200 minutes), the local response, target response and phase error of each meta-atom in the initial design (Fig. 7a) and optimized design (Fig. 7b) are calculated and plotted. This optimization process not only improved the focusing efficiency, but also fixed the focal length shifting problem caused by accumulating phase errors in the initial design (Fig. 7c). The average phase error of the optimized design is 14.6°, largely reduced compared to the 65.3° phase error of the initial design, which increased the electric field intensity at the focal spot from 6.5 V/m to 8.2 V/m.

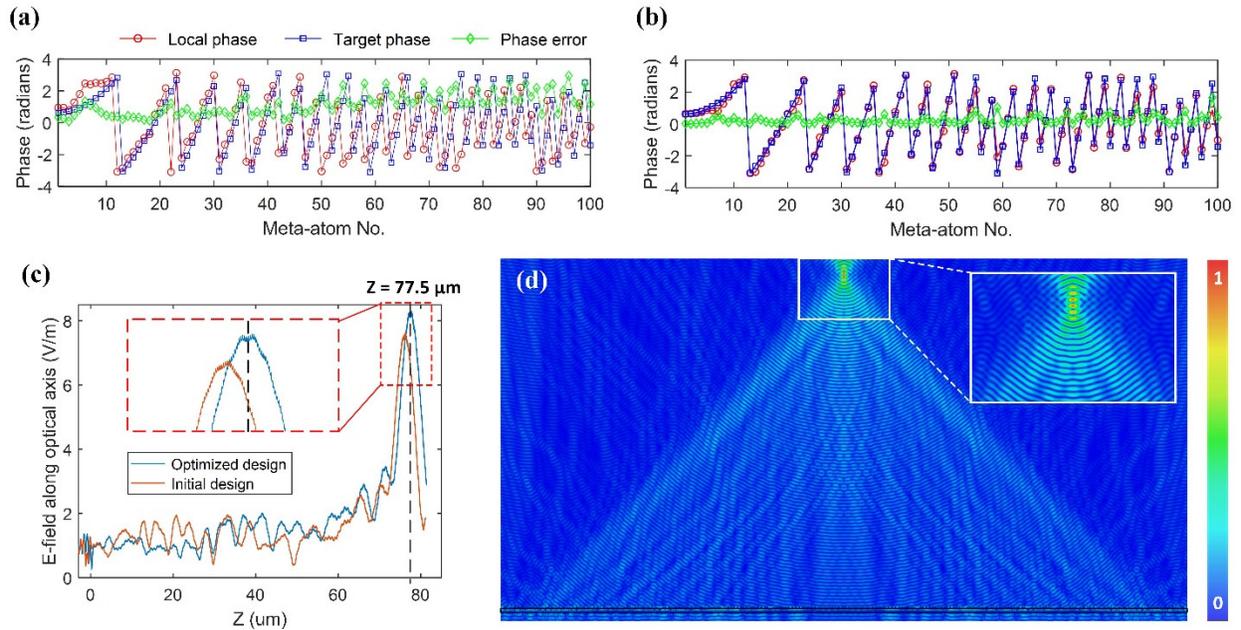

**Fig. 7. Optimization of a larger metalens composed of 200 meta-atoms.** Target phase (blue), local phase (red) and phase error (green) of each meta-atom in **(a)** the initial design and **(b)** the optimized design. **(c)** Simulated electric field along the optical (Z) axis, with an x-polarized incidence for the initial design (red) and the optimized design (blue). The inset contains a magnified view of the E-field near the focal spot. **(d)** Simulated electric field (real part) of the optimized design and magnified view of the E-field near the focal spot (inset).

**Discussion and conclusion.** In this paper, we have quantified the effects of mutual coupling in metasurface designs and devised a novel way of quickly and accurately predicting the perturbed response of a specific meta-atom after being surrounded by different neighbors through a deep learning approach, which can be used to inversely optimize the design's performance. Although the presented PNN and optimizations are based on a simple rectangular-shaped meta-atom design built with a commonly adopted material (polysilicon) in the near infrared range (1550 nm), this approach is not limited to metasurfaces with specific materials, shapes or working frequencies. In Fig. 8a we extend this mutual coupling analysis to another metasurface platform made by patterning a 1-µm-thick film of high-index dielectric material (n = 5) placed on a low-index dielectric substrate (n = 1.4). The periodicity is set to be 2.8 × 2.8 µm$^2$, and the spectrum of interest is 5.45 µm. In this case, to introduce more degrees of freedom, a 2D cross section of each meta-atom is defined by a complex image with resolution of 64 by 64 pixels. High index meta-atoms confine EM fields more strongly than low index meta-atoms and thus are less prone to mutual coupling (Fig. 8b). Therefore, we only consider the influence of 2 neighbors on each side during the data collection process, which provides the same level of accuracy as the previous setup in Fig. 2. Similarly, 10,000 groups of data were collected to assemble the training dataset, and the network architecture was slightly modified to accommodate the new input dimension (320 by 64). When the training was completed, the average MSE was 2e-5 for the training set and 6e-5 for the test set, respectively (Fig. 9a). Six results were randomly selected from the test dataset to show the network's accuracy (Fig. 9b).

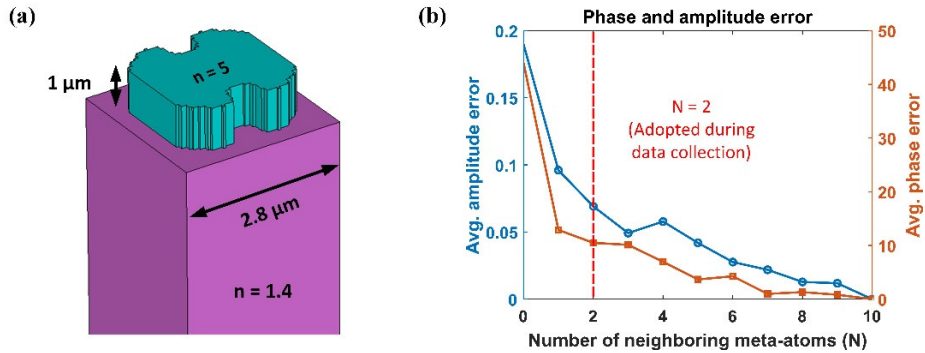

**Fig. 8. Data collection setup. (a)** Schematic of a freeform meta-atom composed of a high-index material sitting on top of a low-index substrate. **(b)** The amplitude and phase MAE for a specified number of non-identical neighboring meta-atoms on each side, N. N = 0 represents the results derived with periodic boundaries. N = 2 was adopted during the data collection process.

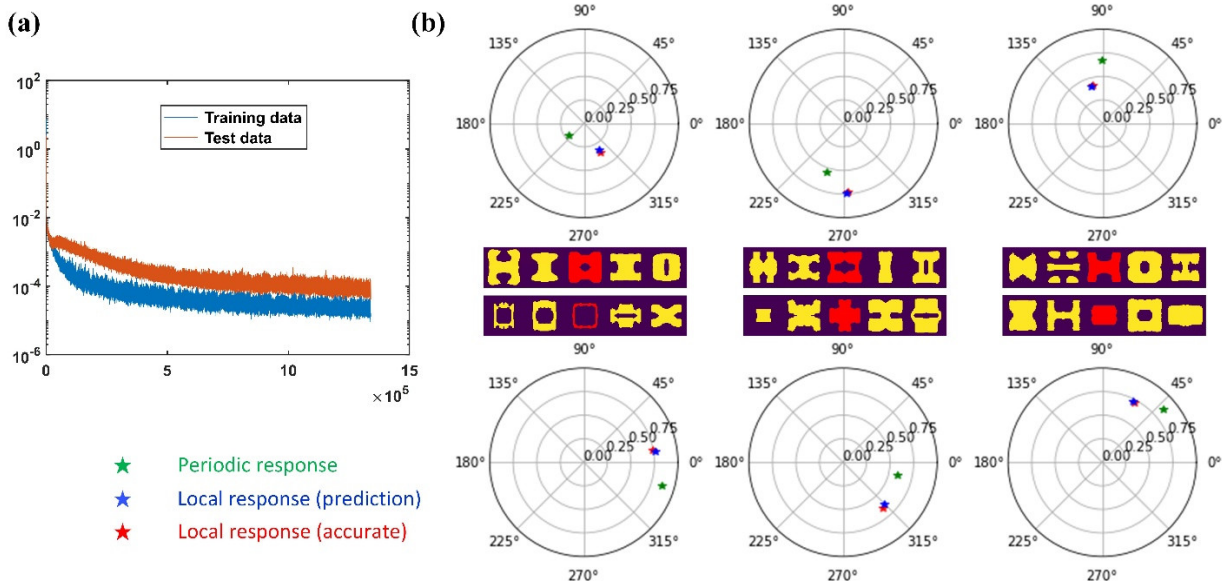

**Fig. 9. Learning curve and results. (a)** Average MSE of the training set (red curve) and test set (blue curve) during the training process. **(b)** Six groups of results randomly selected from the test set are demonstrated.

Besides applications in metasurface optimizations problems, the proposed deep learning approach provides an efficient way to explore the mutual coupling effects between different structures. For example, utilizing this well-trained PNN, we are able to visualize the negative correlations between the metasurface's performance and the severity of the mutual coupling effects. To achieve this, it's necessary to generate and compare various metasurface designs composed of individual meta-atoms with identical EM responses but different shapes, in which case the Generated Adversarial Network (GAN) provides a promising solution to this design task. Here, 400 meta-atoms were generated with a fully-trained GAN [22] that's capable of designing meta-atoms based on phase and amplitude targets. Among these 400 meta-atom designs, each group of 100 meta-atoms is created with the same amplitude target of 0.9 and different phase targets of 45, 135, 225 and 315 degrees, respectively (Fig. 10a). Subsequently, one meta-atom was randomly selected from each set of 100 geometries to assemble a beam deflector consisting of 4 meta-atoms (specified by red lines in Fig. 10a-b). By repeating this process, a total number of 1,000 beam deflectors were created and simulated with a full wave simulation tool to calculate their efficiencies and local responses. The relationship between the efficiency of each deflector and the average MSE of all meta-atoms inside are plotted in Fig. 10b. Although the periodic responses of these meta-atoms are almost identical, their corresponding local responses can be very different when they are placed among different neighbors in a large array (Fig. 10c) due to the mutual coupling effect. To better visualize the relationship between local responses and metasurfaces' performance, we plotted the E-field and local responses of two selected deflectors (Fig. 10d) from these 1,000 designs. The high efficiency (86.3%) designs shown on the right clearly have higher average amplitude and precise 90° phase shifts between adjacent meta-atoms as compared to their low efficiency (51.6%) counterparts shown on the left. The efficiency of each deflector among the 1,000 designs ranges from only 50% to almost 90%, indicating that these high-index freeform mid-infrared metasurfaces also suffer from mutual coupling effects. Moreover, the obvious negative correlation between the error and the overall efficiency demonstrates the feasibility of using the local responses to inversely optimize the performance of a wider variety of metasurfaces.

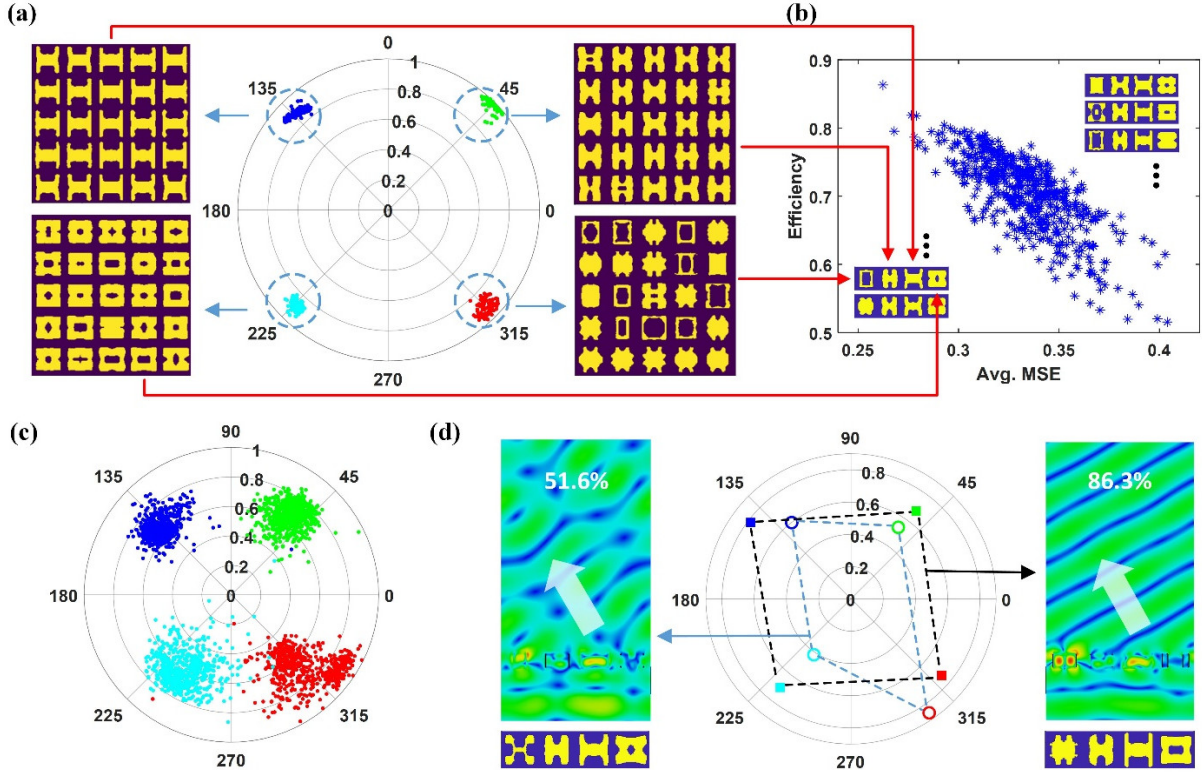

**Fig. 10. Extension to freeform meta-atoms built with high refractive index material. (a)** Periodic phase and amplitude responses of 400 meta-atom designs generated with a fully-trained GAN model. Each group of 100 meta-atoms are created with the phase target: 45° (green), 135° (blue), 225° (cyan) and 315° (red), respectively. Several examples selected from each group are shown as insets. **(b)** MSE versus efficiency of 1,000 beam deflectors composed of 4 meta-atoms that were randomly selected with one from each group of meta-atoms in (a). For demonstration, top views of several deflectors are included as insets. **(c)** Corresponding local responses of all the meta-atoms in the beam deflectors. **(d)** The simulated electric field of two deflectors selected from (b) and the local response of each meta-atom in these two designs.

A major advantage of this deep learning approach is time efficiency. Once fully-trained, the PNN calculates the local responses of target meta-atoms on a one-time calculation basis. As a result, when combined with different optimization algorithms, the PNN can evaluate the performance of the new designs instantly, without the time-consuming fullwave analyses. In conventional meta-optic optimization problems, the time-consuming fullwave validations have been the bottleneck of the whole design process. For example, one fullwave simulation of the metalens in Fig. 7d takes over 30 minutes, which makes the iterative optimization of such a large structure nearly impossible. In contrast, the local responses and the objective function of the metalens can be calculated in seconds, which can largely accelerate the adjoint optimization process. Moreover, parallel optimization algorithms and parallel computing can be introduced to further improve the time efficiency.

A major challenge of this adjoint optimization approach is that the local responses of the meta-atoms cannot always meet the design goals. For example, the optimized beam deflector design in Fig. 5b only achieved 68.8% efficiency, which is still far from the theoretical limit. However, we believe that this issue can be addressed by introducing more design degrees of freedom. For example, compared to the simple rectangular-shaped meta-atom design, the deflector assembled with freeform shaped meta-atoms in Fig. 10d has the potential to achieve much higher efficiency (86%). Meanwhile, more-sophisticated global optimization approaches[15, 34] such as evolution strategies[16, 27, 35, 36] will be needed to deal with the massive design degrees of freedom brought by the freeform-shaped meta-atoms.

To conclude, we have proposed a deep learning network that accounts for mutual coupling effects to efficiently predict the local responses of target meta-atoms. The fully-trained network takes the dimension of a target meta-atom and its neighbors as input and generates its accurate local response in milliseconds. We have demonstrated the network's capability through the optimization of a beam deflector and a metalens with greatly improved performance. Furthermore, we have unveiled and visualized the correlation between the devices' performance and local response errors using the well-trained network. We envision that this deep learning approach will lead to significant improvements in efficiency for large metasurface designs and to metasurface designs surpassing conventional optical components in many applications, including miniaturized optics, holography, and optical information processing.

Acknowledgement

The work is funded under Defense Advanced Research Projects Agency Defense Sciences Office (DSO) Program: EXTREME Optics and Imaging (EXTREME) under Agreement No. HR00111720029.